\documentclass{optica-article}

\journal{opticajournal} 
\usepackage{bm}
\usepackage{
    lineno,
    textcomp,
    float,
    }
\articletype{Research Article}

\newcommand{\vect}[1]{\mathbf{#1}}
\newcommand{\vectrho}{\bm{\rho}}
\newcommand{\appenref}[1]{Appendix A}
\newcommand{\secref}[1]{Sec.~\ref{#1}}
\newcommand{\eqnref}[1]{Eq.~(\ref{#1})}
\newcommand{\figref}[1]{Fig.~\ref{#1}}

\newcommand{\citeasnoun}[1]{Ref.~\citenum{#1}}
\newcommand{\citeasnouns}[2]{Refs.~\citenum{#1}~\&~\citenum{#2}}

\newcommand{\maxwell}[0]{\nabla\times\frac{1}{\mu}\nabla\times\vect{E}-\omega_m^2\vect{\varepsilon}(\vectrho)\vect{E}=\ -j\omega_m\vect{J}}

\begin{document}

\title{Fabrication Tolerant Multi-Layer Integrated Photonic Topology Optimization}

\author{Michael~J.~Probst\authormark{1}, Arjun~Khurana\authormark{1}, Joel~B.~Slaby\authormark{1}, Alec~M.~Hammond\authormark{1,2}, and Stephen~E.~Ralph\authormark{1,*}}

\address{\authormark{1}School of Electrical and Computer Engineering, Georgia Institute of Technology, Atlanta, GA 30308, USA\\
\authormark{2}Now at Meta, 1 Hacker Way, Menlo Park, CA 94025, USA\\ 

\email{\authormark{*}stephen.ralph@ece.gatech.edu}}

\begin{abstract*}Optimal multi-layer device design requires consideration of fabrication uncertainties associated with inter-layer alignment and conformal layering. We present layer-restricted topology optimization (TO), a novel technique which mitigates the effects of unwanted conformal layering for multi-layer structures and enables TO in multi-etch material platforms. We explore several approaches to achieve this result compatible with density-based TO projection techniques and geometric constraints. Then, we present a robust TO formulation to design devices resilient to inter-layer misalignment. The novel constraint and robust formulation are demonstrated in 2D grating couplers and a 3D polarization rotator.

\end{abstract*}

\section{Introduction}
Photonic topology optimization \cite{sigmund_overview,couplers_inverse,carbide_inverse,kim2023automated,hammond_phase} is a design technique which parameterizes a design region into millions of voxels representing unique degrees of freedom and optimizes these voxels to produce freeform devices often with nonintuitive geometries, unprecedented performances, and record small sizes. Multiple layers and etch steps provide the optimizer with more flexibility, but additional fabrication capabilities contain nuainces which have yet to be considered in a TO framework and significantly impact device performance.

\begin{figure}[H]
\centering\includegraphics{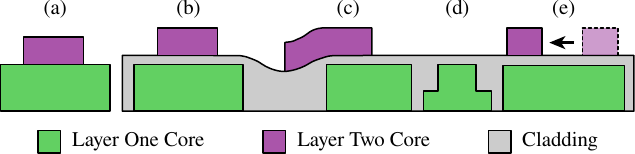}
\caption{Multi-layer integrated photonics processes stack several design layers, either directly (a) or spaced by an inter-layer cladding film (b). This cladding allows the second layer to be placed in regions not physically supported by the bottom layer, but unwanted conformal layering causes sagging on the upper layer (c, \secref{sec:Restrict}). Other processes include multiple etch steps of different depths on the same layer, creating distinct sublayers (d, \secref{sec:Restrict}) which fundamentally cannot support overhangs. Furthermore, the layers or sublayers can become misaligned during fabrication (e, \secref{sec:Misalign}). These two effects invariably degrade device performance, and capturing them in a manner compatible with a TO workflow is challenging and implemented for the first time in this work.}
\label{fig:challenges}
\end{figure}

Intuitively, a simple way to increase performance bounds is to increase the dimensionality of the search space which can be accomplished by increasing the size of the design \cite{liu2018very}. However, increasing the design area footprint may be undesirable with respect to PIC density specifications, and may have minimal impact on performance. Additionally, certain devices benefit from breaking z-symmetry to achieve optimal performance such as polarization-converting structures\cite{Khurana_Inverse-Designed_Photonic_2023,song2023fully}, which require asymmetry to perform mode conversion, and out-of-plane grating couplers, which leverage asymmetry to minimize substrate emission\cite{HammondGrating, cheng2020grating}.  Thus, vertical dimensionality increases are often employed, which are available in several fabrication contexts and may be broken into two major categories:
\begin{enumerate}
        \item \textbf{Multi-layer processing}, in which several layers of (potentially different) materials are deposited and etched (\figref{fig:challenges}a,b).
        \item \textbf{Multi-etch processing}, in which multiple stages of etching are performed on a single layer (\figref{fig:challenges}d).
\end{enumerate} 

We identify two challenges unique to multi-layer and multi-etch processes. The first relates to the fact that in multi-layer processes (fig 1a,b), layer planarization may not be perfectly uniform\cite{weigel2023design}, leading to conformal layering of subsequent depositions and consequent inconsistency between the simulated and fabricated geometry (\figref{fig:challenges}c)\cite{girardi2023multilayer}. A straightforward approach is to prohibit the upper core material voxels from existing over regions of cladding in the lower design layer. Importantly, this approach also enables TO for multi-etch processes, which necessarily prohibit ``overhangs'' of the upper etch steps (\figref{fig:challenges}d). To formulate this approach in a way amenable to a TO paradigm, we propose two distinct solutions: 1) a \textit{projection-based} approach and 2) a nonlinear \textit{constraint-based} approach. The projection-based approach represents a deviation from the conventional constraint-based approaches which are common in photonic inverse design, and instead draws from the rich field of TO for additive manufacturing (e.g., 3D printing)\cite{gaynor2014topology, gaynor2016topology,johnson2018three}. We emphasize that each of these methods is readily integrated with the existing TO pipeline, explore key differences in both approaches, and demonstrate their ability to produce robust, high-performance designs.

The second major challenge presented by multi-layer and multi-etch processes relates to inter-layer/etch misalignment, in which the spatial relationship of two adjacent layers/etches is distorted (\figref{fig:challenges}e). Devices which are tolerant to this misalignment can alleviate associated performance degradations\cite{shang2015low}. To this end, we present a novel implementation of robust TO\cite{wang2011robust,wang2019robust,schevenels2011robust} realized by simultaneously optimizing over layer shifts in each available transverse dimension. Robustness to inter-layer misalignment represents an important step to ruggedizing photonic integrated circuits for first time right design which had previously been absent from the TO literature. 

To illustrate TO solutions to each of the unique challenges posed by multi-layer and multi-etch processes we explore two exemplar structures: out-of-plane grating couplers and polarization rotators. By physical principle and empirical evidence, each of these devices benefits from variable-height waveguide geometries to obtain optimal performance, making them ideal to showcase the layer restrictions. We use throughout this work a 220~nm silicon-on insulator (SOI) waveguide platform with a 160~nm poly-crystalline silicon (poly-Si) overlay, and a relatively conservative minimum length-scale feature size of 100~nm. Of course, our approaches are readily generalizable to any standard material platform and layer stack. Furthermore, throughout this work, we treat multi-etch processes as multi-layer processes, where each etch depth is considered as a separate layer, and use the unified ``layer'' nomenclature to refer both to etch depths and proper layers.  To begin, in \secref{sec:TO}, we give an overview of photonic topology optimization. Next, in \secref{sec:Restrict}, we detail the projection-based and constraint-based TO augmentations enabling multi-etch optimization and mitigating the negative effects of conformal layering. Continuing in \secref{sec:Misalign}, we describe the formulation enabling robustness to inter-layer misalignment. Finally, in \secref{sec:Rotate}, we optimize a fully 3D polarization rotator as a worked example.

\section{Photonic topology optimization overview}\label{sec:TO}
In this section, we outline our previous approach to photonics TO, as summarized in \citeasnouns{hammond_constraints}{hammond2022high} and explain specific enhancements used in this work. We formulate conventional broadband photonic TO as a multi-objective minimization problem over $N$ objective functions, subject to Maxwell's equations \cite{maxwell1865viii} at $M$ frequency points and $K$ constraint functions:
\begin{equation}\label{eq:obj_old}
    \begin{matrix}
         & \min\limits_{\vectrho}\big\{f_n(\vect{E})\big\} & n\in\left\{1,...,N\right\} \\
         \text{s.t.} & \maxwell & m\in\left\{1,...,M\right\} \\
         &0\leq\vectrho\leq1 & \\
         & g_k(\vectrho)\leq0 & k\in\{1,...,K\}
    \end{matrix}
\end{equation}
where $f_1,...,f_N$ are performance objectives dependent on the electric field $\vect{E}$, $\vectrho$ is the design variables which determine the electric permittivity $\varepsilon$ of the device, $\vect{J}$ is the source current density, and $g_k$ are constraints on the design variables. Often, $N$ is a multiple of $M$ corresponding to multiple objectives (e.g., target transmission, reflection, \& crosstalk) at each frequency.  Optimizing the design variables to achieve a topology with excellent performance that also satisfies the constraint functions requires gradually enforcing the optimization constraints. To prevent the optimizer from settling into an unsatisfactory local minimum, we recast the optimization problem in a novel form:
\begin{equation}\label{eq:obj}
    \begin{matrix}
         & \min\limits_{\vectrho}\left[\displaystyle\sum\limits_{n=1}^N\bar{g}_n\big(f_n(\vect{E}),\vect{q}_n\big)+\sum\limits_{k=1}^K\bar{g}_{N+k}\big(g_k(\vectrho),\vect{q}_{N+k}\big)\right] & n\in\left\{1,...,N\right\},\ k\in\{1,...,K\}\\
        \text{s.t.} & \maxwell & m\in\left\{1,...,M\right\}\\
        &0\leq\vectrho\leq1&\
    \end{matrix}
\end{equation}
where $\bar{g}_i,\ i\in\{1,...,N+K\}$ are differentiable spline-based scaling functions \cite{messac1996physical} applied to both the objectives and the constraints and $\vect{q}$ are user-defined lists that separate the evaluation of each objective/constraint into categories. The returns of each objective/constraint, which originally had been over different scales and units, are now optimized on a unified, unitless, scale generated from the user-defined performance bounds. Thus, the problem of hyperparameter optimization to satisfy both the objectives and constraints sufficiently is greatly simplified, and the possibility of the optimizer minimizing one objective/constraint at the expense of another is mitigated. 

We project the latent (unfiltered) design parameters $\vectrho$ to the permittivity distribution $\varepsilon$ using a filtering-threshold parameterization (also called the three-field scheme) \cite{sigmund_overview}. The filtering implicitly enforces a minimum feature size and the thresholding encourages binarization which is ultimately necessary for fabrication. To employ gradient based optimization techniques through the parameterization, both operations must be differentiable. The linear filter acting on the latent design variables is applied with the convolution
\begin{equation}
    \widetilde{\vectrho}=w(\vect{x})*\vectrho,
\end{equation}
where $*$ denotes 1D or 2D convolution and $\widetilde\vectrho$ are the filtered design variables. $w(\vect{x})$ is a filter described by:
\begin{equation}\label{eq:conic}
    w(\vect{x})=
    \begin{cases} 
      \frac{1}{a}\left(1-\frac{|\vect{x} - \vect{x}_0|}{R}\right), & \vect{x} \in \mathcal{N} \\
        0, & \vect{x} \notin \mathcal{N}
   \end{cases}.
\end{equation}
Here, $\mathcal{N}$ is the ``neighborhood'' of a parameter, or the set of surrounding voxels which inform how it transforms under the filter. In the 1D case, $\mathcal{N}$ is a line segment of length $2R$ centered at $\vect{x}_0$ and $w(\vect{x})$ is a triangular filter. In the 2D case, $\mathcal{N}$ is a disk with center $\vect{x}_0$ and diameter $2R$, and $w(\vect{x})$ is a conic filter. $R$ corresponds to the largest minimum feature size of the foundry process and $a$ normalizes $w$ such that $\smallint d\vect{x} w(\vect{x})=1$. The filtered design parameters then undergo thresholding via a nonlinear filter which encourages binarization. We employ the differentiable modified hyperbolic tangent projection:
\begin{equation}\label{eq:threshold}
    \bar{\vectrho}=\frac{\rm tanh \left(\beta\eta\right)+\rm tanh \left(\beta\left(\vect{{\widetilde{\vectrho}}}-\eta\right)\right)}{\rm tanh\left(\beta\eta\right)+\rm tanh\left(\beta\left(1-\eta\right)\right)},
\end{equation}
where $\bar{\vectrho}$ are the thresholded design parameter and $\beta$ and $\eta$ are user-controlled parameters that describe the steepness and center of the projection respectively. The value of $\beta$ denotes different design epochs where the design smoothly progresses from grayscale optimization to quasi-binary optimization. The thresholded parameters are linearly interpolated between the solid $(\varepsilon_\text{max})$ and void $(\varepsilon_\text{min})$ phases of the relative permittivity:
\begin{equation}\label{eq:permittivity}
    \vect{\varepsilon}_r(\bar{\vectrho})=\vect{\varepsilon}_\text{min}+\bar\vectrho\big(\vect{\varepsilon}_\text{max}-\vect{\varepsilon}_\text{min}\big).
\end{equation}
We perform the forward and adjoint Maxwell solves using the open-source finite difference time domain (FDTD) simulator Meep\cite{meep} and implement the adjoint variable method (AVM)\cite{Strang07,steven_adjoint} using its built-in solver\cite{hammond2022high} to compute the gradient of the FOM with respect to all the design parameters at all design wavelengths. The gradient is backpropagated through the parameterization with a vector-Jacobian product (vJp) implemented by the open-source software package Autograd\cite{maclaurin2015autograd}.  The latent design variables are optimized using the globally convergent method of moving asymptotes (GCMMA)\cite{svanberg_mma}, provided by the open-source nonlinear optimization package NLopt\cite{nlopt}.

\section{Layer restrictions}\label{sec:Restrict}
We demonstrate two approaches to producing topologies that bypass the fabrication challenges of multi-etch and multi-layer design: a projection-based approach and a constraint-based approach. In each of the following subsections, the latent design regions are initialized as two arrays of 1-dimensional design parameters, $\vectrho_1$ and $\vectrho_2$. In the projection-based approach (\secref{sec:projection}), the \textit{permittivities} describing the top and bottom layers become more elaborate functions of these parameters, and in the constraint-based approach (\secref{sec:constraint}), the permittivities follow the conventional parameterization (\secref{sec:TO}).  Both approaches offer unique advantages (\secref{sec:comparison}) and are useful in the TO toolkit.

\subsection{Approach one: projection-based approach} \label{sec:projection}
\begin{figure}
\centering\includegraphics[width=5.25in]{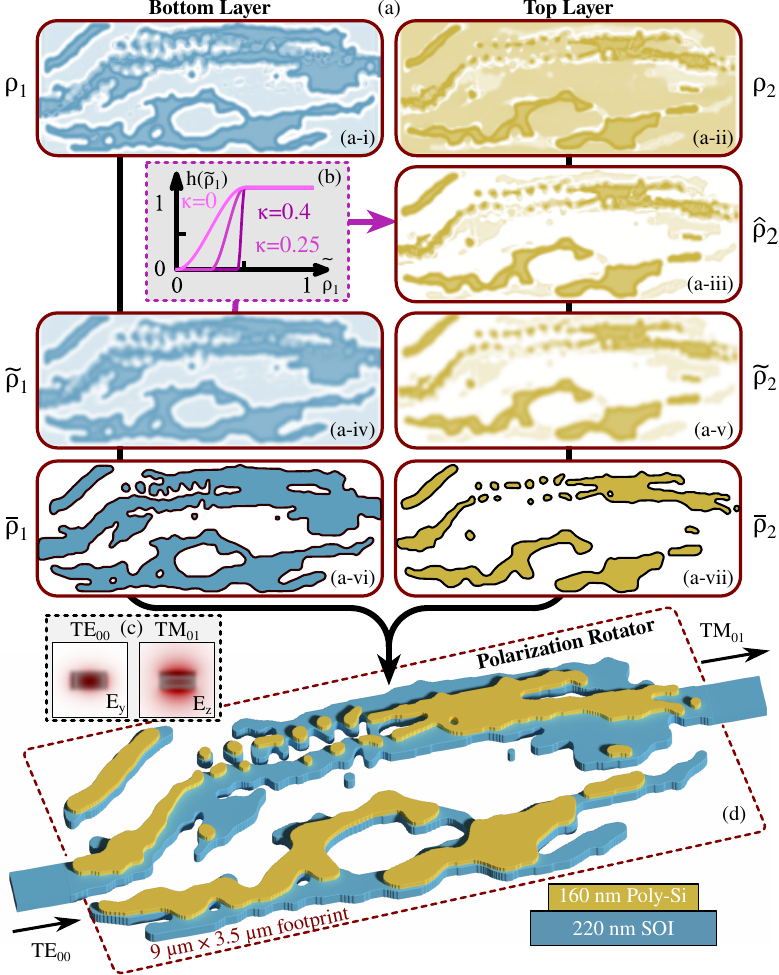}
\caption{A TO parameterization for two layer designs enforces the top layer is physically supported by the bottom layer (a). The layer restriction is implemented as an enhancement to the conventional filter-threshold  scheme\cite{zhou_geometric}. Specifically, the filtered design parameters of the bottom layer (a-iv) are scaled by a nonlinear curve (b) with steepness parameter $\kappa$ and multiply the unfiltered design parameters of the top layer (a-ii). This is called the overlap filter, and it enforces the top layer only exist over filled regions of the bottom layer. Both layers are thresholded and extruded into a 3D structure for simulation. This parameterization is demonstrated on a two-layer TE$_{00}$ to TM$_{01}$ (c) polarization rotator (d) in 220~nm SOI (blue) with a 160~nm poly-Si overlay (yellow).}
\label{fig:mapping}
\end{figure}

The projection-based approach projects the top layer design parameters into the subspace defined by the solid phase of the bottom layer design parameters. Therefore, anywhere the bottom parameters are in the void phase, the top parameters must also be in the void phase. To implement this, prior to the filter-threshold operation on the top layer, we define the overlapped design parameters, $\hat\vectrho_2$, such that:
\begin{equation}
    \hat{\vectrho}_2=h\big(\widetilde\vectrho_1\big)\times{\vectrho}_2,
\end{equation}
where $\times$ denotes element-wise multiplication and $h$ is a differentiable nonlinear filter acting on the bottom filtered design parameters (\figref{fig:mapping}b). We define the overlap filter $h$ to be:
\begin{equation}\label{eq:overlap_filter}
    h(\widetilde\vectrho_1,\kappa)=\begin{cases}
        0 & \widetilde\vectrho_1<\kappa\\
        -2(\tau(\widetilde\vectrho_1-\kappa))^3+3(\tau(\widetilde\vectrho_1-\kappa))^2 & \kappa\leq\widetilde\vectrho_1\leq0.5 \\
        1, & \widetilde\vectrho_1>0.5
    \end{cases},
\end{equation}
where $\kappa$ represents the lower rise point of the overlap filter and $\tau$ is a convenience factor defined as:
\begin{equation}
        \begin{matrix}
            & \tau=\cfrac{1}{0.5-\kappa}, & 0\leq\kappa<0.5.        
        \end{matrix}
\end{equation}
The purpose of $h$ is to scale the bottom filtered design parameters and ultimately control how quickly the layer restriction is enforced during the optimization. This is realized with the independent parameter $\kappa$, which controls the steepness of the overlap filter (\figref{fig:mapping}b). Specifically, voxels in the void phase on the bottom $(\widetilde\vectrho_1\approx0)$ force the overlapped top parameters located directly above them to the void phase, whereas voxels that are solid phase $(\widetilde\vectrho_1\approx1)$ have no effect on the top design parameters. However, during the initial stages of the optimization, the design parameters are highly uncertain $(\widetilde\vectrho_1\approx0.5)$, so the parameter $\kappa$ controls the extent to which these parameters affect the top parameters. Low values of $\kappa$ expand the region around $\widetilde\vectrho_1=0.5$ where the bottom design parameters have no effect on the top layer, and high values of $\kappa$ decrease this region. This control is necessary because the overlap filter significantly changes the topology of the top design parameters (and consequently the device performance), requiring careful enforcement of the restriction to maintain excellent performance. Furthermore, the overlap filter operates on the filtered design parameters $\widetilde\vectrho_1$ and not on the thresholded design parameters $\bar\vectrho_1$ to decouple the layer restriction process from the device binarization process. Because the layer restriction significantly impacts the topology, it is enforced early in the optimization (i.e. before the constraints and binarization) so that the optimizer can discover a new local minimum with the layer restrictions in place. Lastly, while $\kappa$ can never be exactly 0.5 during optimization because $h$ converges to the nondifferentiable (and discontinuous) Heaviside step function, $\kappa=0.5$ is useful for final performance evaluation, when the gradient is not calculated, and was applied for all performance objectives shown in this work.

Since the AVM returns the gradient of the FOM with respect to the thresholded design parameters, we backpropagate through the parameterization function to compute the gradient of the FOM with respect to the latent design parameters. Applying the chain rule to the modified parameterization function yields:
\begin{equation}
    \label{eq:drho_1}
    \cfrac{d\text{FOM}}{d\vectrho_1}=\left(\cfrac{\partial\text{FOM}}{\partial\bar{\vectrho}_1}\ \cfrac{d\bar{\vectrho}_1}{d\widetilde{\vectrho}_1}+\cfrac{\partial\text{FOM}}{\partial\bar{\vectrho}_2}\ \cfrac{d\bar{\vectrho}_2}{d\widetilde{\vectrho}_2}\ \cfrac{d\widetilde{\vectrho}_2}{\partial\hat{\vectrho}_2}\cfrac{\partial\hat{\vectrho}_2}{d\widetilde{\vectrho}_1}\right) \cfrac{d\widetilde{\vectrho}_1}{d\vectrho_1}
\end{equation}
\begin{equation}
    \label{eq:drho_2}
    \cfrac{d\text{FOM}}{d\vectrho_2}=\cfrac{d\text{FOM}}{d\bar{\vectrho}_2}\ \cfrac{d\bar{\vectrho}_2}{d\widetilde{\vectrho}_2}\ \cfrac{d\widetilde{\vectrho}_2}{\partial\hat{\vectrho}_2} \cfrac{\partial\hat{\vectrho}_2}{d\vectrho_2}.
\end{equation}
A consequence of the overlap filter is the thresholded design parameters on the top layer are a function of the latent design parameters from \textit{both} the top and bottom layers, and the vJp must be constructed properly to compute the gradient of the FOM with respect to the latent design parameters of both layers\cite{hammond2022high}.

\citeasnoun{zhou_geometric} introduced two differentiable constraints which conform the optimized design to the minimum linewidth and linespacing constraints prescribed in a foundry's design rulebook and also aid in the binarization of the thresholded design parameters. These constraints rely on the specific interplay between the latent, filtered, and thresholded design parameters, so any modification to the parameterization must be applied \textit{before} the filter-threshold operation. For this reason, the overlap filter is applied to the \textit{latent} top design parameters, $\vectrho_2$. In principle, however, the filter-thresholding operation may not preserve satisfaction of the layer restrictions at the boundaries between solid and void phases. This is especially true of the filtering operation, which may shift boundaries as a consequence of averaging voxels with high degrees of uncertainty ($\vectrho_2\approx0.5$) together with neighboring voxels of low uncertainty ($\vectrho_2\approx0,1$). However, we note the lengthscale constraints create abrupt boundaries between the solid and void regions of the latent design parameters, as illustrated by the distinct blue/white boundries of $\vectrho_1$ and yellow/white boundaries of $\hat\vectrho_2$ in \figref{fig:mapping}a. Therefore, the locations of these boundaries are invariant under the filter-threshold operation, meaning the overlap filter can be applied to the latent design parameters and the thresholded design parameters will also conform to the layer requirement.

\subsection{Approach two: constraint-based approach}\label{sec:constraint}
The second approach is a constraint-based approach which removes overhanging regions in the top layer. Constraint based approaches are the conventional technique toward enforcing any topological requirement in the photonic TO ecosystem \cite{piggott2017fabrication,hammond_constraints,zhang2021topological,zhou_geometric}. The optimization problem (\eqnref{eq:obj}) easily enables additional constraints as all constraint returns are unified on the same scale by the spline-based scaling functions\cite{messac1996physical}, without the need to add any machinery to the design process. We define the constraint function:
\begin{equation}
    g_{LC}=\cfrac{1}{N}\sum\limits_{i=1}^N\text{max}\big\{\bar\rho_{i,2}-\bar\rho_{i,1},0\big\},
\end{equation}
which identifies areas where the top parameters are in the solid phase over areas where the bottom parameters are in the void phase. The gradient of this constraint encourages the bottom parameters to the solid phase and the top parameters to the void phase, conforming the topology to the layer restriction. Heuristically, the layer constraint places the largest limitations on the allowed topologies and has the largest impact on the performance compared to other DRC constraints. Therefore, we typically weight the spline-based scaling bounds, $\vect{q}$ \cite{messac1996physical}, to satisfy the layer constraint in tandem with the performance objectives early in the optimization (low values of $\beta$) an apply the DRC constraints during the final $\beta$ epoch. While the ``max'' operation introduces a non-differentiability in the evaluation of the constraint function, enforcing the constraint on the thresholded design parameters cleverly avoids it. Because $\bar\vectrho$ is quasi-binary, the only constraint violations are $\bar\vectrho_2-\bar\vectrho_1\approx1$, which is far from the non-differentiability.

\subsection{Comparison of the two approaches}\label{sec:comparison}
The two implementations of the layer restriction contain significant similarities and differences, yet both are useful in the TO design flow. Firstly, both approaches are compatible with the traditional constraints, parameterization techniques, and robust formulations found in the TO literature. Next, an important functionality for optimization-based design is the ability to tune hyperparameters to guide the optimizer to a satisfactory local minimum. This tuning ability is realized in the projection-based approach by the parameter $\kappa$, which describes the strictness of the spatial overlap filter (\eqnref{eq:overlap_filter}) and in the constraint-based approach by the designer's choice of the spline-based scaling bounds $\vect{q}$ (\eqnref{eq:obj}) \cite{messac1996physical}, which weight the importance of each constraint. As mentioned in \secref{sec:constraint}, the constraint-based approach is trivial to integrate into the TO pipeline because \eqnref{eq:obj} permits an arbitrary number of constraints. The projection-based approach, albeit more difficult to integrate into the TO pipeline due to the modifications in the backpropagation (Eqs. \eqref{eq:drho_1} \& \eqref{eq:drho_2}), is more transparent to the optimizer than the layer constraint. Previous literature from TO for additive manufacturing supports the conclusion that projection-based approaches are simpler for the optimizer than constraint-based approaches\cite{gaynor2014topology}, and this is consistent with our observations that devices designed with the projection-based approach require less tuning. In contrast, an inherent advantage of the constraint-based approach is the ability to retain or delete the gradient of either layer which may aid the optimization depending on the specific layer stackup. For example consider a device where the top layer does not have a large effect on the fields (e.g., the top layer has low index, is thin, or is separated by a thick inter-layer cladding). Then, it may be beneficial optimize the bottom parameters solely for performance and DRC constraints, and only retain the layer constraint gradient with respect to the top parameters, which intuitively have a smaller impact on the performance.

To demonstrate the efficacy of the two approaches for the layer restriction with a simple yet illustrative design, we designed three 1D multi-layer fiber grating couplers. The first grating featured no layering restrictions, the second utilized the modified projection, and the third included the layer constraint. The optimization problem is represented in \eqnref{eq:obj} with a FOM given by:
\begin{equation}
    f_n(\vect{E})=10\times\text{log}_{10}\left(\cfrac{|\alpha_0^+|^2}{P_n}\right),
\end{equation}
where $\alpha_0^+$ is the overlap of the simulated fields and the fundamental mode of the integrated waveguide pointed away from the grating and $P_n$ is power emitted by the source which normalizes the overlap integral. The mode overlap for the $m^\text{th}$ mode (in either propagation direction is given by:
\begin{equation}
    \label{eq:overlap}
    \alpha_m^\pm=\int\limits_A\left[\vect{E}^*(r)\times \vect{H}_m^\pm(r)+\vect{E}_m^\pm\times\vect{H}^*(r)\right]\cdot d\vect{A},
\end{equation}
where ``$+$'' denotes the forward propagating modes, ``$-$'' the backward modes, $d\vect{A}$ is the differential cross-sectional area of the waveguide pointing normal to the direction of the waveguide, $\vect{E}(r)$ and $\vect{H}(r)$ are the amplitudes of the simulated electric and magnetic fields at optimization wavelengths, and $\vect{E}^\pm_m(r)$ and $\vect{H}^\pm_m(r)$ are the mode profiles for the $m^\text{th}$ mode in the direction of the mode overlap. Specifically, we defined one FOM at each $\lambda = 1540\text{ nm, }1550\text{ nm, } 1560$ nm (i.e. $M=N=3$). The gratings were optimized over a $10\ \upmu$m long design region to couple the TE$_{0}$ integrated waveguide mode into standard single mode fiber (SMF) positioned $1\ \upmu$m above the SOI with a launch angle of $15^\circ$. The SMF mode was modeled by a Gaussian source with $10.4\ \upmu$m beam waist\cite{marchetti2017high}. The simulations were performed at resolution $50\ \text{voxels}/\upmu$m, and the optimization was performed with a resolution of $100\ \text{voxels}/\upmu$m. We performed lengthscale constraints\cite{zhou_geometric} on all three designs to enforce a minimum feature size of 100~nm.

\begin{figure}[ht!]
    \centering
    \includegraphics{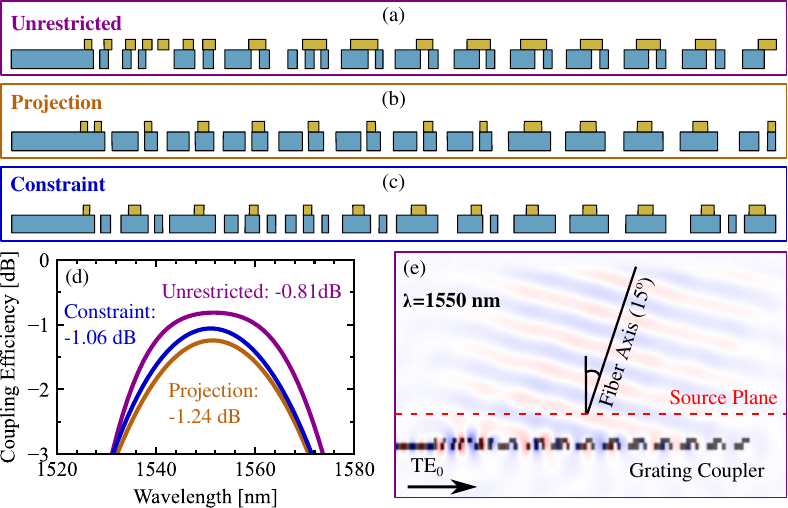}
    \caption{Two-layer grating couplers were optimized with unrestricted layers (a), layers projected through the overlapping filter (b, \secref{sec:projection}), and layers restricted by a constraint function (c, \secref{sec:constraint}). In the optimization without layer restrictions, the optimizer favored a topology with overhanging regions which enabled -0.81~dB coupling efficiency. The layer-restricted gratings still achieved competitive coupling efficiencies of -1.24~dB (projection) and -1.06~dB (constraint) without the use of overhanging regions (d). The field plot of the layer-unrestricted grating is well confined to the fundamental fiber mode (e) at launch angle $15^\circ$.}
    \label{fig:gratings_constraint}
\end{figure}

The grating with unrestricted layers (\figref{fig:gratings_constraint}a) achieved the lowest loss of -0.81~dB (\figref{fig:gratings_constraint}d), whereas the gratings optimized with the projection-based approach (\figref{fig:gratings_constraint}b) and constraint based approach (\figref{fig:gratings_constraint}c) achieved losses of -1.24~dB and -1.06~dB, respectively (\figref{fig:gratings_constraint}d). From the field plot of the constraint-based design excited from the waveguide (\figref{fig:gratings_constraint}e), the gratings launch at the appropriate angle with a small amount of power directed downward and into higher order diffraction angles. Both methods of implementing the layer restriction reduced the search space compared to the unrestricted design, which clearly utilized overhanging regions to achieve $\sim$0.3~dB higher performance. While the limited search space predictably contributed to the lower performance, it should be noted that the final performances of these optimizations are local minima; this result is not indicative of the performance discrepancy between restricted and restricted layers for any optimization. Rather, the modified projection or constraint should be applied to processes that either require (multi-etch) or would benefit from (unwanted conformal layering) the layer restrictions to improve the \textit{measured} performance. Lastly, we note these results were achieved using standard a fabrication platform without techniques such as optimizing cladding or layer thicknesses. With these additional optimizations and a less conservative feature size, we expect to achieve gratings that adhere to the layer restrictions with sub-0.5~dB coupling efficiency.

\section{Layer misalignment}\label{sec:Misalign}
We formulate a novel TO algorithm enabling multi-layer structures which are robust to inter-layer misalignment. Inter-layer misalignment \cite{fleming1999three} may be on the order of tens of nanometers and invariably causes degraded device performance\cite{HammondGrating}. This is especially true of TO structures which have traversed a narrow optimization path toward a relatively sharp local minimum. Our methodology leverages the well-known technique of robust TO, which utilizes the inherently parallelizable nature of TO to simultaneously optimize several transformed versions of a nominal design, resulting in a final design with higher tolerance to those transformations \cite{hammond_constraints}. In the case of modelling layer misalignment, this transform is accomplished by convolving the upper design parameters, $\vectrho_2$, with a delayed delta kernel:
\begin{equation}
	\vectrho_2' = \vectrho_2 \ast \delta(x - \Delta),
\end{equation}
where $\Delta$ is a parameter describing the degree of inter-layer misalignment. 

Integration of the inter-layer misalignment algorithm requires compatibility with all parts of the TO pipeline, including FOM definition, geometric lengthscale constraints, and layer restrictions. Since the performance of the nominal and layer-shifted designs are evaluated in independent simulations, and FOM (and gradient) for each design is averaged in the optimization problem (\eqnref{eq:obj}) integration with the FOM definition is straightforward. Additionally, geometric lengthscale constraints are only applied to the nominal design, since the inter-layer misalignment does not change the topology of either layer, only the spatial relationship between them. Thus, we need only consider the interaction between inter-layer misalignment and our newly defined layer restrictions. 

We employ the constraint-based approach in this work due to its straightforward implementation: the layer restriction constraint is applied in parallel to each of the nominal and shifted designs, such that if an inter-layer misalignment results in violations of the layer restriction constraint, it is penalized accordingly. We note that in principle the layer-restriction constraint need not be applied to the nominal design, since the satisfaction of the constraint by the shifted designs necessarily implies that the nominal design also satisfies (assuming that $\Delta$ is small compared to the minimum lengthscale). However, we still apply the constraint to all designs for the sake of algorithmic simplicity.

By contrast, the projection-based approach would require significantly more machinery. To illustrate this, note that the shift operation and parameterization function commute, so we can consider without loss of generality the ordering of first applying the shift, then applying the parameterization. Now, consider the effect of a shifted upper layer resulting in overhang above a void region in the lower layer. The nonlinear projection $h$ in the parameterization function will cause this region to become void. However, in a real fabrication context, the presence of inter-layer misalignment does not result in the conversion of unetched regions to etched regions. Thus, this projection would result in non-physical simulations. To account for this, an additional step could be included which enforces a so-called ``run length'', i.e., the lower design region core material must extend sufficiently far past the upper design region core material that the shift does not result in overhangs. In principle, this could be accomplished by applying an erosion to the bottom layer before the overlap filter, either via a morphological transform \cite{svanberg2013density,hammond_constraints} or by modifying the $\eta$ parameter of the thresholding function\cite{wang2011robust}. However, both of these approaches realize the erosion on the thresholded design parameters, and we recall that the new nonlinear projection $h$ requires access to the filtered (but not thresholded) lower design parameters $\widetilde\vectrho_1$. Thus, these methods are not available. We consider a novel method of enforcing the run length, which modifies the filtered design parameters:
\begin{equation}\label{eq:neweta}
	\widetilde\vectrho_1' = \widetilde\vectrho_1 + \Delta \eta,\quad \Delta \eta \equiv \eta_\text{mod} - \eta
\end{equation} 
which has a similar effect to modifying the $\eta$ parameter for small $\Delta \eta$, but in a way that is transparent with respect to $h$. 

This can be thought of as shifting the filtered design parameters by some amount so that if they were thresholded with \eqnref{eq:threshold} and $\eta=0.5$, the result would be the same as if the unsubtracted filtered parameters were thresholded with $\eta$ defined by \eqnref{eq:neweta}. As discussed in \cite{hammond_constraints}, the thresholding approach is problematic because it is difficult to calculate the dilation/erosion caused by decreasing/increasing $\eta$ and the topology is not necessarily consistent through the operation (i.e. features appear and disappear because of the filtering/thresholding), especially if $\eta$ is changed. However, we submit that this method is reliable in our design scheme for two reasons. 1) The geometric lengthscale constraint maximize the dynamic range of the boundaries between the solid and void phases of the unfiltered design parameters, meaning $\eta$ has a predictable effect on the amount of dilation and erosion in the thresholded design parameters. 2) The misalignment applied in this work (40~nm) is much smaller than the filter size; a small change in $\eta$ is necessary to realize the dilation. This precludes the possibility of features disappearing during the erosion.  

\begin{figure}[ht!]
    \centering
    \includegraphics{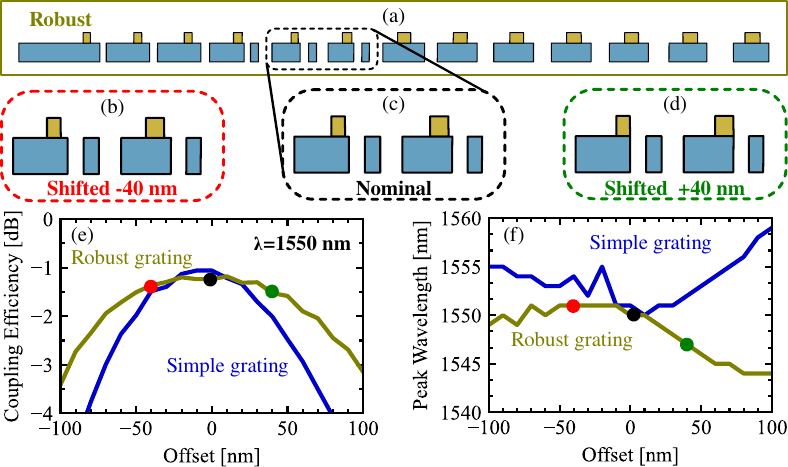}
    \caption{A two layer grating coupler (a) is optimized for robustness to inter-layer misalignment. The nominal (unshifted) design (c) and designs featuring an inter-layer shift of $\pm40$ nm (b,d) were simultaneously optimized for performance and satisfaction of the layer constraint (\secref{sec:constraint}). The gratings are evaluated at $\lambda=1550$~nm swept over inter-layer offset. With 0~nm offset between the layers, the robust and simple grating exhibited similar coupling, and as the inter-layer offset increased, the robust grating outperformed the simple grating (e). Furthermore, the peak coupling wavelength of the robust grating shifted less than the peak coupling wavelength of the simple grating on account of the offset (f).}
    \label{fig:robust_grating}
\end{figure}

To demonstrate the robust TO formulation, we compare the 1D grating coupler designed with the layer constraint with a robust grating coupler designed with the layer constraint. Simulations approximate the width of the grating coupler as infinite, meaning any inter-layer misalignment across the width of the grating will have negligible impact on performance. Thus, we need only consider misalignments down the length of the grating (in the direction of the integrated waveguide propagation). Using the same hyperparameter set as in Sections 2 and 3, we optimized a robust layer conforming grating coupler assuming inter-layer misalignment of $\pm40$ nm. \figref{fig:robust_grating}a illustrates the final permittivity of the grating. Note that the nominal (\figref{fig:robust_grating}c) and shifted (\figref{fig:robust_grating}b,d) designs adhere to the layer constraint because it was applied on all three fields concurrently. Furthermore, the performance of the robust device at $\lambda=1550$ nm varies only 0.16~dB across $\pm40$ nm of inter-layer misalignment (\figref{fig:robust_grating}e) compared to the non-robust design which varies 0.9~dB. While both gratings exhibit peak coupling wavelength close to $\lambda=1550$~nm with less than $40\ \upmu$m offset, the peak coupling wavelength of robust grating remained close to $\lambda=1550$~nm for offsets up to 100~nm while the peak coupling ratio of the non-robust showed a red shift (\figref{fig:robust_grating}f).

\section{Numerical example: polarization rotator}\label{sec:Rotate}
\begin{figure}[H]
    \centering
    \includegraphics{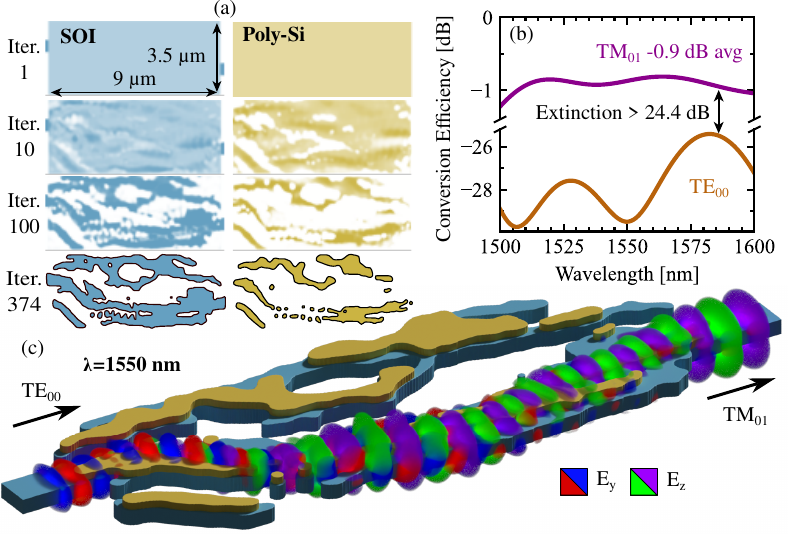}
    \caption{A TE$_{00}$ to TM$_{01}$ polarization rotator was designed with the modified projection function (\secref{sec:projection}) to enforce the layer restriction. The device evolution at iterations 1, 10, 100, and 374 are shown (a) for both the SOI layer (blue) and the poly-Si (yellow) overlay. The rotator demonstrated exceptionally broadband performance with an average conversion efficiency of -0.9~dB from 1500 nm to 1600~nm. Though extinction was not explicitly maximized, it is greater than 24.4~dB across the entire design band (b). The devices is excited with a 1550~nm continuous wave at the TE input, and the $E_y$ (red/blue) and $E_z$ (green/purple) fields are shown (c).}
    \label{fig:rotator_performance}
\end{figure}

We optimized a 3D two layer TE$_{00}$ to TM$_{01}$ polarization rotator using the modified projection function described in \secref{sec:Restrict} to implement the layer restriction. We satisfied \eqnref{eq:obj} with an FOM defined by
\begin{equation}
    f_n(\vect{E})=10\times\log_{10}\left(\ \left|\cfrac{\alpha_1^+}{\alpha_0^+}\right|\ \right),
\end{equation}
where $\alpha_1^+$ is the mode overlap of the simulated fields and the second waveguide mode (from \eqnref{eq:overlap}) which corresponds to the TM$_{01}$ mode. We optimized one objective at ten wavelengths evenly spaced between $\lambda=1500$~nm and $\lambda=1600$~nm. The polarization rotator was optimized over a $9\ \upmu$m by $3.5\ \upmu$m design region connected at the short edges to two 500~nm wide waveguides offset by $1\ \upmu$m (\figref{fig:rotator_performance}a). A simulation resolution of $30$ voxels$/\upmu$m and a design resolution of 60~voxels~$/\upmu$m were used, resulting in 228,000 design parameters over the two layers. We optimized the device over three epochs using $\beta=8,\ 16,\ 32$ (\eqnref{eq:threshold}) and $\kappa=0.48$ (\eqnref{eq:overlap}). During the final epoch, we concurrently enforced geometric linewidth and linespacing constraints to restrict the minimum feature size to 100~nm. We terminated the optimization when the constraints were satisfied and performance stopped improving after 374 iterations (\figref{fig:rotator_performance}a). The optimization was distributed on two Intel Xeon Gold~6226 2.7~GHz nodes (48 cores) provided by the Partnership for an Advanced Computing Environment (PACE) at the Georgia Institute of Technology\cite{PACE}. Each iteration (two Maxwell solves) completed in $\sim$6 minutes, for total optimization time of less than 40 hours.

To characterize the final performance, we completely binarized the design with a Heaviside step function centered at $\widetilde\vectrho=0.5$ (since the threshold function in \eqnref{eq:threshold} cannot fully binarize and remain differentiable) and simulated 100 wavelengths between 1500~nm and 1600~nm, verifying the device performance between the optimization wavelengths. The rotator achieved TE$_{00}$ to TM$_{01}$ conversion efficiency of $-1.04\pm0.24$~dB across the design band, and, though not specifically optimized, the TE$_{00}$ to TE$_{00}$ extinction was greater than 24.4~dB (\figref{fig:rotator_performance}b). The rotator topology deviates significantly from conventionally designed polarization rotators, which enact the rotation over several hundred microns using offset waveguides of varying materials, thicknesses, and widths. From the field plot in \figref{fig:rotator_performance}c the conversion between the TE$_{00}$ and TM$_{01}$ modes happens gradually throughout the length of the device, and the optimizer favored using a curved waveguide-like topology to enact the conversion in just 9~$\upmu$m. The efficiency, therefore, could be improved by a longer design region and/or larger waveguide offset. Furthermore, large features of the design seemingly have no impact on the electric field, given that the electric field is confined to one side of the topology. During the grayscale phase of the optimization, these regions assisted in the polarization conversion, but as the optimizer settled into a minimum, the resultant structure did not have field near those features. However, since they do not negatively affect the performance, the optimizer did not remove them \cite{bobbyto}. For our previous topology optimized polarization rotator, the experimental performance was within 1~dB of simulation \cite{Khurana_Inverse-Designed_Photonic_2023}; moveover, since the layer constraint reduces experimental degradation due to unwanted conformal layering, we expect this device to perform to within 0.5~dB of simulation\cite{probst_nitride}.
\section{Conclusion}\label{sec:conclusion}
We have demonstrated several TO augmentations that combat the fabrication challenges of multi-layer photonic design. Our techniques are compatible with standard TO parameterizations and conventional constraint functions. We displayed two novel approaches to restricting the top layer over the bottom layer: a projection which projects the top layer into the subspace defined by the bottom layer and a constraint which penalizes ``overhangs'' on the top layer. Then, we implemented a novel robust formulation for inter-layer misalignment and designed a robust 2D grating coupler. Our work utilized a standard 220~nm SOI process but is extensible to arbitrary layer thickness and material platforms. With additional optimizations, we are confident our methodology can produce experimentally validated record low-loss passive routing structures and IO devices. 

Future work includes implementing a more elaborate projection based on cross-section data of the fabricated layers that captures the extent of nonplanarization and projects the design parameters into an appropriate 3D structure that represents the fabricated device topology. Furthermore, the projection techniques presented here can be extended to fabrication processes with nonsquare profiles such as blazed and angled sidewalls (e.g. thin-film lithium niobate). Lastly, designs could be optimized concurrently over additional uncertainties such as varying layer thicknesses and refractive index to ruggedize the designs. 
\begin{backmatter}
\bmsection{Funding}
This material is based upon work supported in part by the National Science Foundation (NSF) Center ``EPICA'' under Grant No.1 2052808, \url{https://epica.research.gatech.edu/}. Any opinions, findings, and conclusions or recommendations expressed in this material are those of the author(s) and do not necessarily reflect the  views of the NSF. MJP, AK, JBS, and SER were supported by the Georgia Electronic Design Center of the Georgia Institute of Technology.

\bmsection{Acknowledgments}
This research was supported in part through research cyberinfrastructure resources and services provided by the Partnership for an Advanced Computing Environment (PACE) at the Georgia Institute of Technology, Atlanta, Georgia, USA. This material is based upon work supported in part by the National Science Foundation (NSF) Center ``EPICA'' under Grant No. 2052808, \url{https://epica.research.gatech.edu/}. The authors would like to thank Steven~G.~Johnson for his useful discussions.

\bmsection{Disclosures}
The authors declare no conflicts of interest.

\bmsection{Data availability} Data underlying the results presented in this paper are not publicly available at this time but may be obtained from the authors upon reasonable request.

\end{backmatter}

\bibliography{references}

\end{document}